 \documentclass[prl,preprint]{revtex4-1}

\usepackage{amsmath}  % needed for \tfrac, \bmatrix, etc.
\usepackage{amsfonts} % needed for bold Greek, Fraktur, and blackboard bold
\usepackage{graphicx, subfigure} % needed for figures

\usepackage{color}
\definecolor{r}{rgb}{1,0,0}

\usepackage[normalem]{ulem}

\usepackage{epsfig}
\newcommand{\rs}[1]{{\color{black}  #1}}

\newcommand{\delete}[1]{{}}

\begin{document}

\title{\rs{Secondary flow in ensembles of non-convex granular particles under shear}}

\author{Mahdieh Mohammadi, Dmitry Puzyrev, Torsten Trittel and Ralf Stannarius}

\affiliation{Institute of Physics, Otto von Guericke University, Magdeburg, Germany}

\date{\today}

\begin{abstract}
Studies of granular materials, both theoretical and experimental, are often restricted to convex grain shapes.
We demonstrate that a non-convex grain shape can lead to a qualitatively novel macroscopic dynamics. Spatial crosses
(hexapods) are continuously sheared in a split-bottom container. Thereby, they develop a secondary flow profile that
is completely opposite to that of rod-shaped or lentil-shaped convex grains in the same geometry. The crosses at the surface  migrate
towards the rotation center and sink there, mimicking a `reverse Weissenberg effect'. The observed surface flow field suggests
the existence of a radial outward flow in the depth of the granular bed, thus forming a convection cell. This flow field
is connected with a dimple formed in the rotation center. The effect is strongly dependent on the particle geometry
and the height of the granular bed.
\end{abstract}

\maketitle

%\section{Introduction}

Granular materials had fundamental importance in human civilization since millennia, but still, their dynamical and structural behavior
is much less understood than that of ordinary solids, liquids or gases, and it is often quite counter-intuitive. Ensembles of hard spherical grains have been studied extensively, and important
progress was, and still is, achieved regarding, e.~g., packing \cite{Snoeijer2004,Madani2021,Borzsonyi2016}, shear characteristics and flow \cite{Guazzelli2018,Vo2020,Shaebani2021}, jamming \cite{Majmudar2007,Dauchot2005}, and internal stress distributions \cite{Hashemi2018,Murphy2019,Liu2021}. Recent research has been increasingly dedicated to shape-anisotropic
(e. g. \cite{Donev2004,Borzsonyi2013,Ashour2017a}) and soft grains \cite{Dijksman2013,Hong2017,Ashour2017,Harth2020,Wang2021}.

We focus here on nonconvex particles that exhibit astonishing new features. Research on such particles has been performed only scarcely. A %comprehensive
review of packing problems of particles with various shapes was given by Torquato and Stillinger \cite{Torquato2010}.
Alonso-Marroquin \cite{Alonso-Marroquin2008} developed a simulation tool for 2D nonconvex objects that was extended to 3D by
Galindo-Torres \cite{Galindo2009}.
Az\'ema \cite{Azema2013} numerically investigated stress response and shear strength,
defining and analyzing a `level of convexity'.
Saint-Cyr \cite{Saint-Cyr2011} simulated force chains in 2D, controlling nonconvexity by the choice of special grain geometries.
The packing fraction was shown to grow first and then to decay with nonconvexity. In further simulations,
stickiness was included  \cite{Saint-Cyr2013}. The stress response to cyclic shear was studied
by Athanasiadis \cite{Athanassiadis2014}. Galindo-Torres \cite{Galindo2009} computed the influence of nonconvexity on friction, and Han \cite{Han2021} simulated nonconvex grains flowing down an inclined plane. Sheared ensembles of U-shaped particles in 2D were  simulated by Marschall \cite{Marschall2015}.

\begin{figure}[ht]
\ \hfill
   \includegraphics[width=0.3\columnwidth]{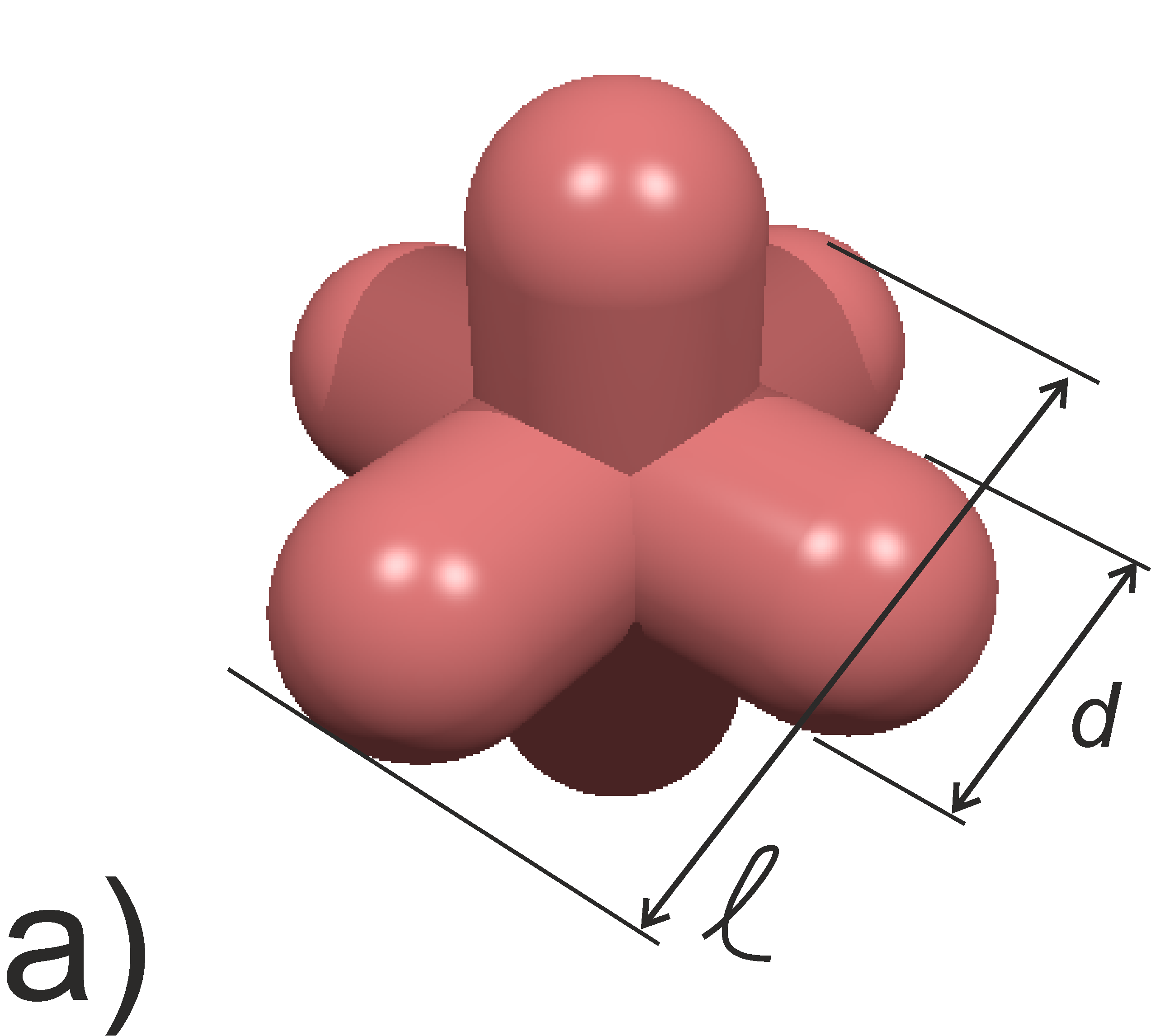}\hfill
   \includegraphics[width=0.5\columnwidth]{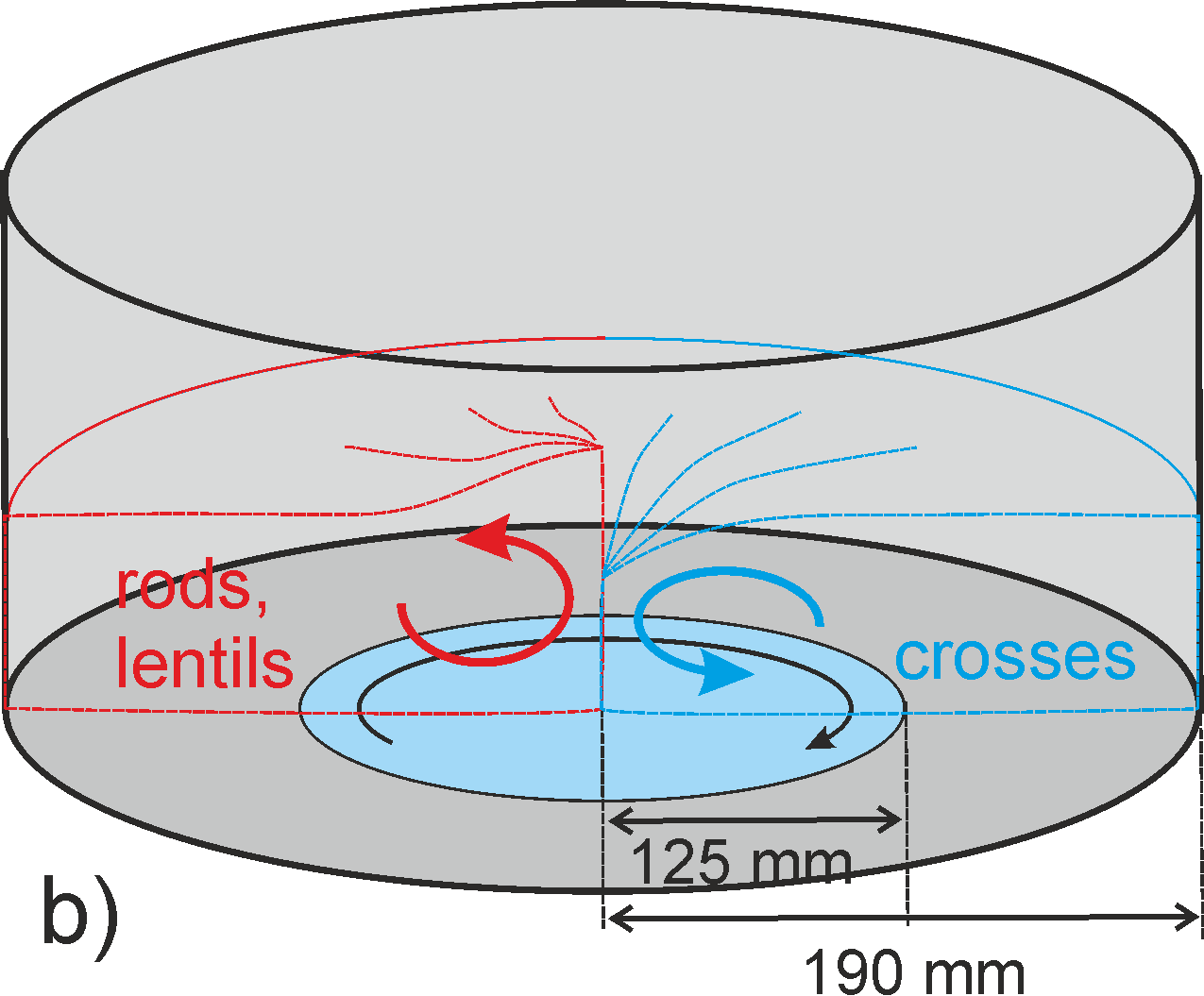}\hfill\
  \caption{\rs{a) Sketch of the geometry of the spatial crosses with $\rho=3$.
  b) Shear geometry. The central bottom
  disk is rotated with a controlled speed. The left (red) and right (blue) dashed lines sketch the central cross sections of the granular
  bed. In red, we sketch earlier observations with elongated cylinders and lentils, (left) contrasted to the present observations for the spatial crosses in blue (right).} Arrows indicate the sense of convection.}
  \label{fig:hexapod}
\end{figure}

Some recent studies dealt with crosses:
Huet \cite{Huet2021} performed   simulations and experiments of the collapse of heaps of 2D crosses.
Their packing in 2D was investigated  theoretically \cite{Atkinson2012,Marschall2020,Meng2020} and experimentally
\cite{Zheng2017,Stannarius2022}.
Spatial crosses, also referred to as hexapods, recently received attention.
Conzelmann \cite{Conzelmann2020} computed their packing and the distribution of contact forces. An experimental method to reveal the
local structure of aggregates of spatial crosses with X-ray scanning was presented by Bar\'es \cite{Bares2017}.
Kuhn described the motion of contacts during slow loading \cite{Kuhn2017}, and
Zhao \cite{Zhao2020} reported shearing of crosses with different aspect ratios $\rho = \ell/d$ (maximum extension
$\ell$ divided by the arm width $d$, see Fig.~\ref{fig:hexapod}a). Ratios between 1 (sphere) and 10 were considered.
The crosses exhibit yield stress when $\rho$ is small ($ < 3.3$). Stiffening sets in at $\rho \approx 6.7$.
In addition to fundamental interest in such particle shapes, stars and hexapods may be of practical value in
granular architecture \cite{Dierichs2015,Dierichs2016,Dierichs2021,Keller2016}.

We shear ensembles of spatial crosses (Fig.~\ref{fig:hexapod}a), employing a split-bottom container   \cite{Fenistein2003,Fenistein2004} with an inner radius of the cylindrical vessel of 190 mm, and a radius $R_0=125$ mm of the rotating bottom disk (sketched in Fig.~\ref{fig:hexapod}b). This disk is coated with a grid structure that prevents the crosses from sliding on the disk surface. The crosses have
an extension $\ell=9.6$~mm and an arm width $d=3.2$~mm
($\rho=3$).
\rs{They were purchased from {\em Yun Nan Yun Tian Hua Co., Ltd}. The particles were produced from polyformaldehyde by injection molding in a custom shaped mold.}
With 50,000 particles available, fill heights $H$ up to 140 mm can be realized.
The bottom disk is rotated by a motor with controlled  angular velocities $\omega_0$ at rates between 3 rpm and 6 rpm. \rs{In this range, we have established that the only influence of the rotation rate is the scaling of the dynamics with the number of rotations.} The top of the granular bed is observed with a stereo camera (Intel D435) that simultaneously captures images of the granular bed and the spatially resolved depth information (1280 px $\times$ 720 px).
The in-plane resolution is 0.33 mm/pixel and the depth resolution was better than 0.5 mm. Typically, we take 10 images per full revolution of the bottom disk.

In a similar setup, an interesting phenomenon was observed earlier: for sheared spherical grains, the granular bed remains practically flat, while for elongated cylindrical grains \cite{Wortel2015,Fischer2016} as well as for flat lentils \cite{Fischer2016}, shear causes the formation of a heap in the rotation center. This heaping is directly connected with a secondary flow in the granular bed: a convection directed radially outward at the bed surface, and towards the center at the bottom.
Figure \ref{fig:hexapod}b in the left half sketches a cross-section
of the granular bed and the convective flow found in earlier experiments with rods, rice grains and lentils.
The heaping effect was found to be particularly efficient when the fill height $H$ of the container was roughly 0.6 times the bottom disk radius $R_0$.
At this height, the material at the top above the bottom disk rotated with about half the disk rotation rate. When the fill level was lower, the top material rotated faster, but the heap formed much slower and with a substantially lower height. When the granular bed was much higher than $0.6R_0$, the material at the surface above the bottom disk rotated slower, the heap also formed much slower, and the heap height as well was considerably reduced \cite{Wortel2015,Fischer2016}. \rs{The absence of heaping and secondary flows for spherical grains (6 mm airsoft bullets) has been confirmed in the present study before the experiments with crosses were performed.}
We anticipate that the experiments with spatial crosses show exactly the opposite phenomenon compared to rods and lentils. This is demonstrated for different fill heights $H$. The observed flow and surface profile are sketched qualitatively in Fig.~\ref{fig:hexapod}b, right.

\begin{figure}[htbp]
\center
      a)\includegraphics[width=0.28\columnwidth]{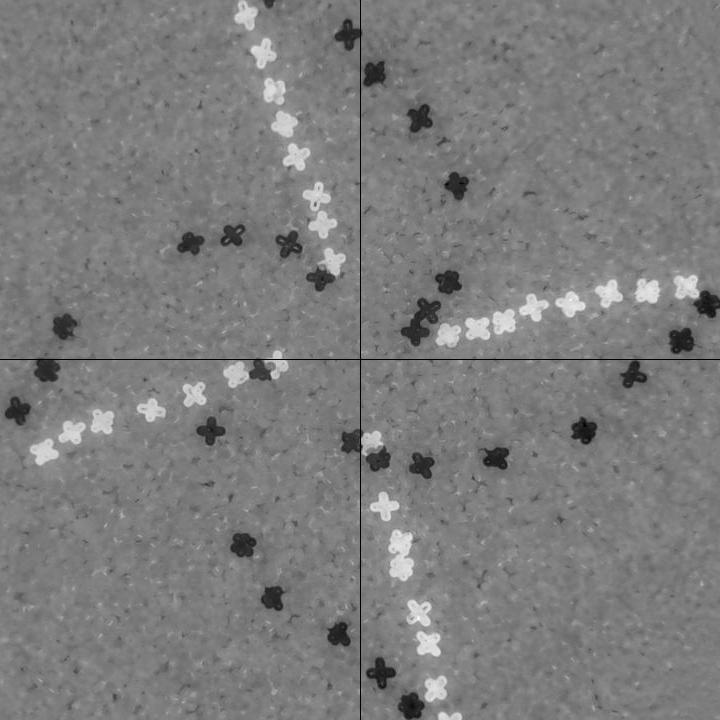}
      b)\includegraphics[width=0.28\columnwidth]{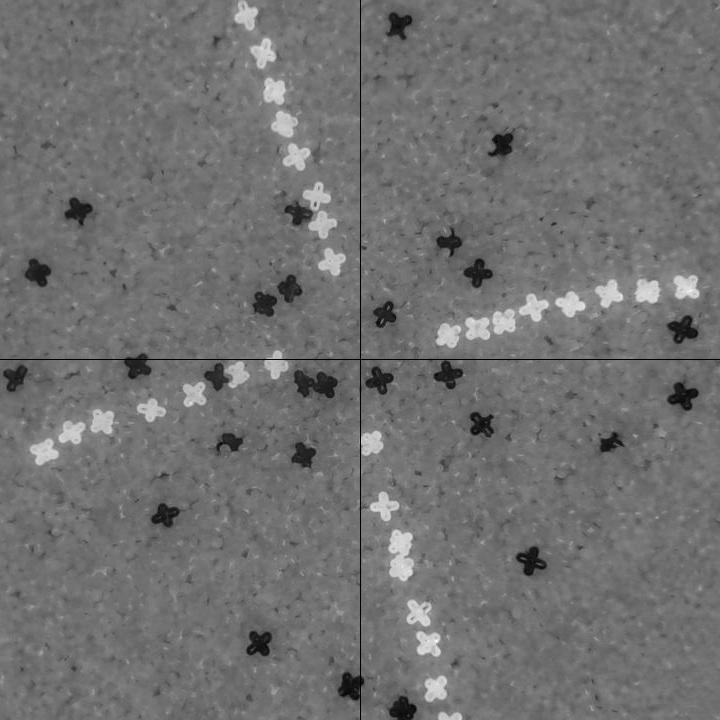}
      c)\includegraphics[width=0.28\columnwidth]{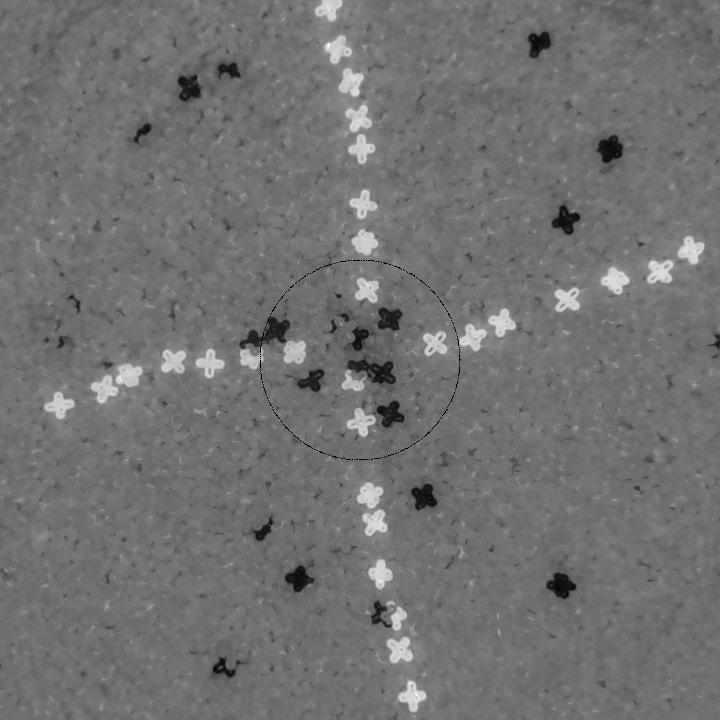}
\caption{Superimposed frames: initial image inverted, markers in white, and images after 5 rotations (a) and 15 rotations (b) with marked crosses in black, $H=100$ mm ($0.8R_0$), $\omega_0= 6$ rpm. c) same construction after 15 rotations for $H= 90$ mm ($0.72 R_0$). Image sizes 24 cm $\times$ 24 cm. The tracer crosses in the central part move inward, crosses in the shear zone and in the outer parts on average maintain their radial positions. The intersection of the fine horizontal and vertical lines marks the rotation center.}
  \label{fig:tracers}
\end{figure}

\rs{In the above-cited earlier studies of non-spherical grains in a similar geometry, the role of the shear zone profile and surface flow in heap formation was clearly established. Therefore,} we first retrieved the tangential flow profile at the granular bed surface to map the shear zone. For that purpose, some black crosses were distributed on top. By tracing them, the local flow field can be probed.
Figure \ref{fig:tracers}a shows the initial image subtracted (markers appear white) from that taken after 5 rotations (black markers) for $H/R_0=0.8$.
In the central part (radius $\approx 5$~cm), the granular surface rotates nearly uniformly, but with only about 2.5 \% of $\omega_0$ (Fig.~\ref{fig:tracers}a). In addition to the rotation, a slow mean radial
flow inward of the order of 1 mm/rotation is evident  (Fig.~\ref{fig:tracers}b). With increasing distance from the center, where the shear zone reaches the surface
(radius between $\approx 5$~cm and $\approx 15$ cm), the surface rotation decreases monotonously, the particles only
migrate diffusively. The inward motion ceases. Beyond the bottom disk edge, there is hardly any observable surface motion.

\begin{figure}[htbp]
\center
      \includegraphics[width=0.7\columnwidth]{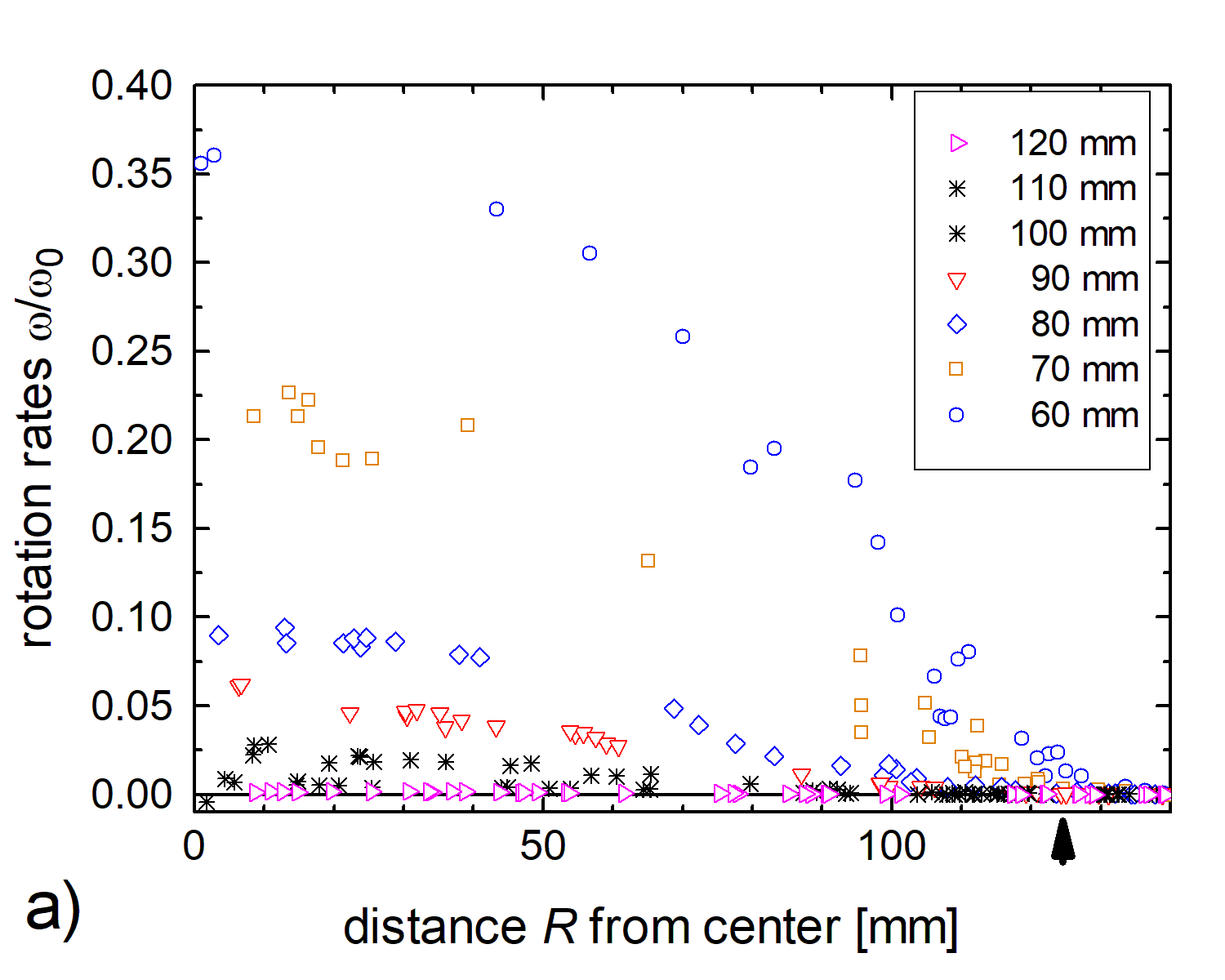}
      \includegraphics[width=0.7\columnwidth]{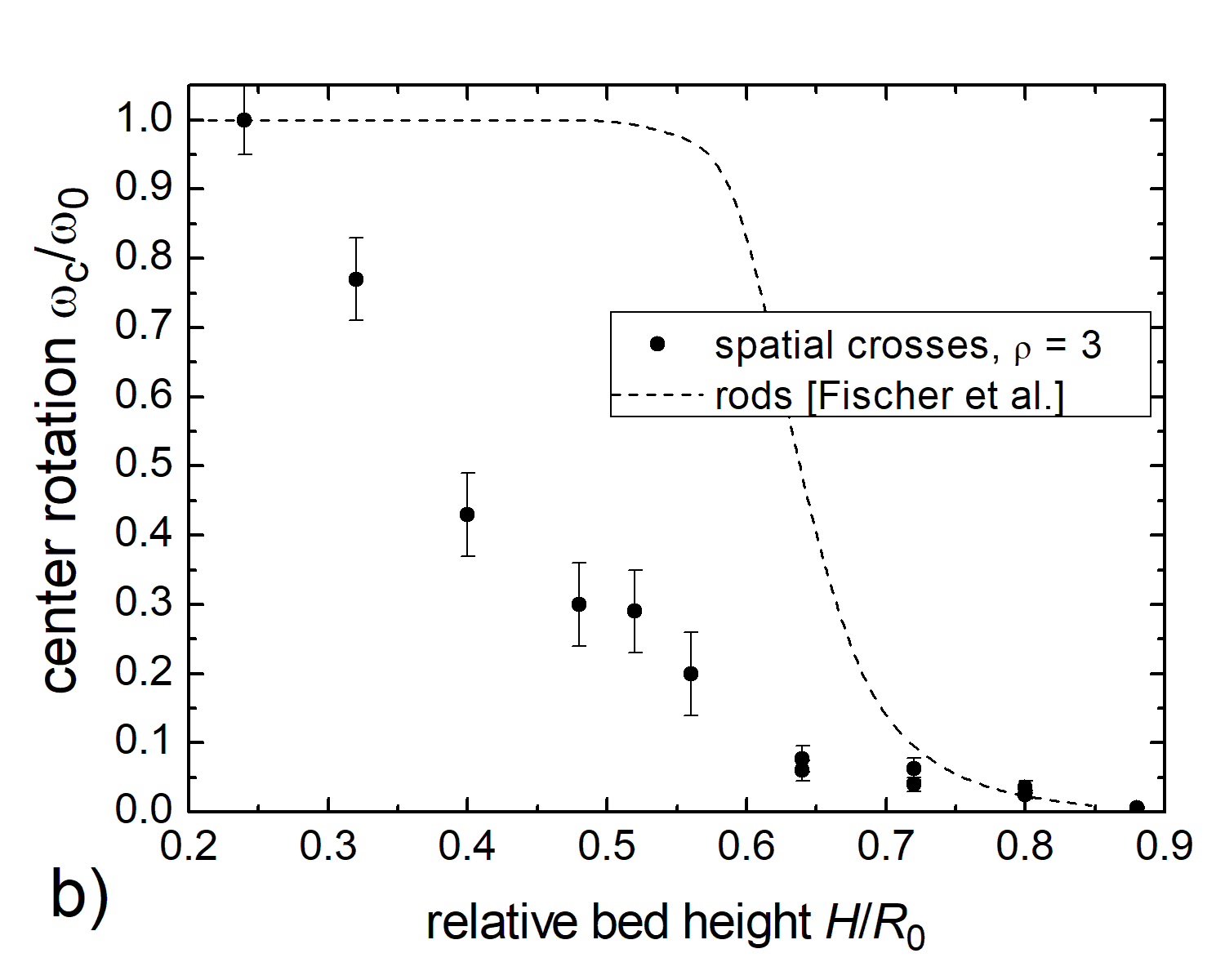}
\caption{a) Surface rotation obtained from of marked tracers in different distances $R$ from the rotation center, given as their angular velocity divided by $\omega_0$. Symbols and colors represent different filling heights (60 to 120 mm). The arrow marks the bottom disk radius. b) rotation of the center for spatial crosses at different fill heights $H$. The dashed line was found for rods \cite{Fischer2016}.}
  \label{fig:tangentialFlow}
\end{figure}

\rs{Because the radial and tangential displacements of the tracers occur at different time scales, we consider both components separately.} Figure \ref{fig:tangentialFlow}a shows the average tangential tracer motion (angular velocities of their orbits around the center) in relation to the bottom disk rotation $\omega_0$. Data were averaged over 8 to 20 disk rotations. As expected, the rotation speed decreases with increasing $H/R_0$,
and with
larger distance $R$ from the center. Figure~\ref{fig:tangentialFlow}b gives the maximum rotation rate $\omega_c$ at the center, scaled with $\omega_0$. Compared to rod-like particles \cite{Fischer2016} the crosses at the surface orbit much slower at comparable bed heights. \delete{A probable reason that waits to be confirmed by X-ray tomography is that friction between the crosses is considerably larger than with the surface structure of the bottom disk. Then, the crosses directly above the disk may start to slide.}

\begin{figure}[htbp]
   \includegraphics[width=0.75\columnwidth]{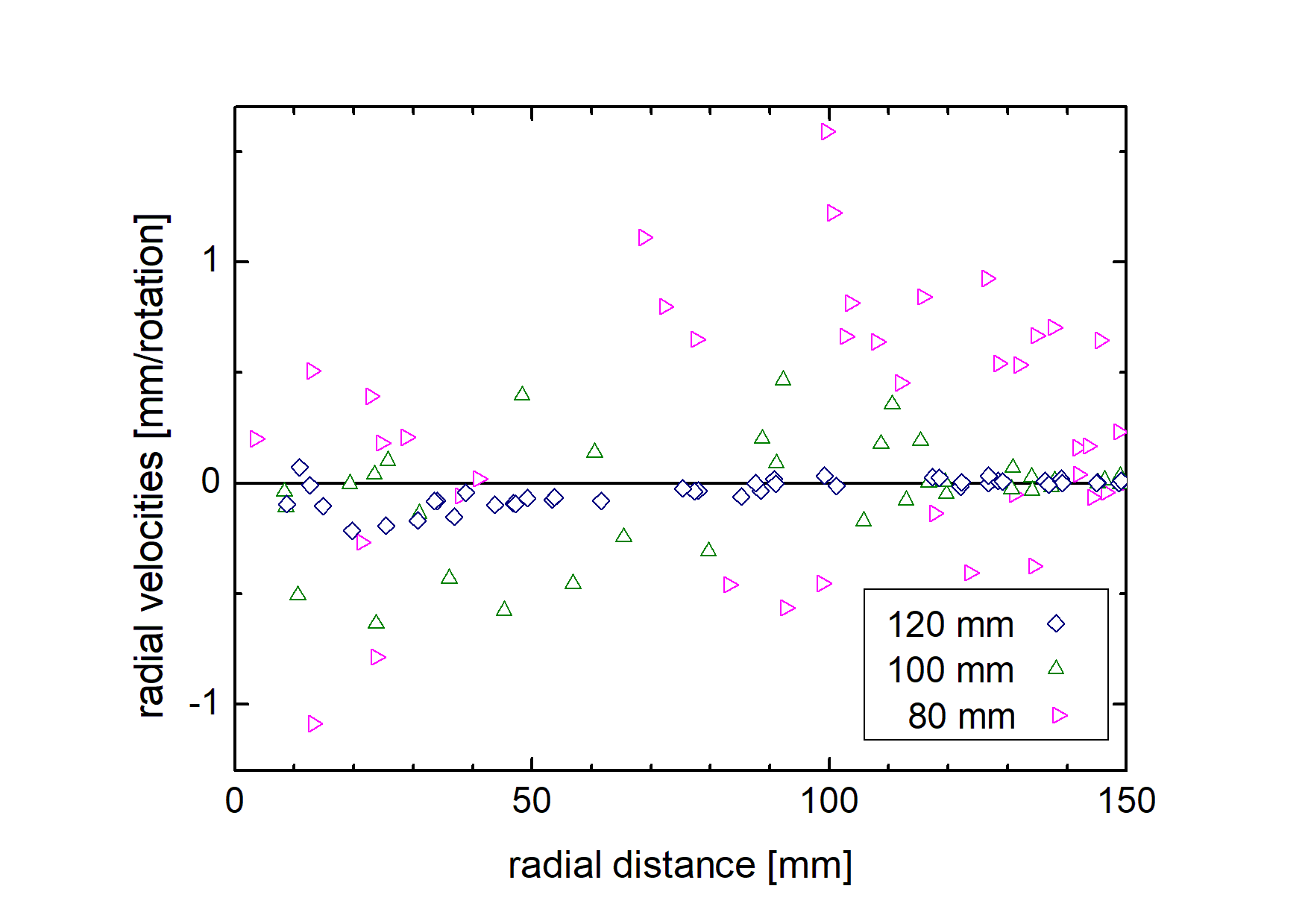}
  \caption{Radial tracer displacements for three different fill heights, $0.96 R_0$, $0.8 R_0$ and $0.64 R_0$}
  \label{fig:radial}
\end{figure}

The radial displacement of the tracers is shown in Fig.~\ref{fig:radial}. It is considerably slower than the tangential motion, and partly obscured by a diffusive
motion of the individual tracers. Again, radial displacements were determined as averages of 8 to 20 rotations. It is seen that the overall radial
flow velocities increase considerably with lower fill heights, while they decrease in higher granular beds. In the central region, there is a clear trend to
negative $v_r$, i.~e. flow towards the rotation center, in the data shown in Fig.~\ref{fig:radial}. In the $H=120$~mm sample, all values except one within a radial distance of about 100 mm ($0.8~R_0$) are negative. They are comparably small, of the order of -0.1 mm/rotation. In the $H=100$~mm bed, there is much faster transport towards the center,
with velocities up to -0.6 mm/rotation. \rs{The superimposed diffusive component is substantially larger.} When the fill height $H$ is further decreased, the diffusive character of radial motion further increases, but there is still a net transport towards the center in the region within $R<0.5 R_0$. With lower $H/R_0$ down to about $0.5$, the diffusive motion becomes prevalent. A net inward flow is observed only in a central area within $R<0.25 R_0$.
\delete{, and net outward flow dominates elsewhere.}

In general, the trend is that at large $H/R_0$, there is less diffusion and a
prevalence of slow directed flow towards the center.  For intermediate fill levels, both the diffusion and directed flow intensify. A separation of a region close to the container center with dominating inward surface flow and an outer region with prevalent outward surface flow is seen. For lower fill heights, the inward flow region shrinks and the diffusive motion and outward flow dominate the movement of particles at the surface.
Eventually, for very low fill heights ($H<0.5 R_0$), the radial transport ceases. Below $H/R_0<0.4$, the center rotates
nearly like a solid block, and both radial flow and diffusion in the central part vanish.

Since we observe a net inward motion of surface particles, at a stationary surface profile (see below), there must be a net transport radially outward in the deeper layers of the granular bed. Thus, it is reasonable to assume a convection as sketched in Fig.~\ref{fig:hexapod}b. The maximum convection
speed is found at fill heights between 0.6 $R_0$ and 0.8 $R_0$.

\begin{figure}[htbp]
  \includegraphics[width=0.26\columnwidth]{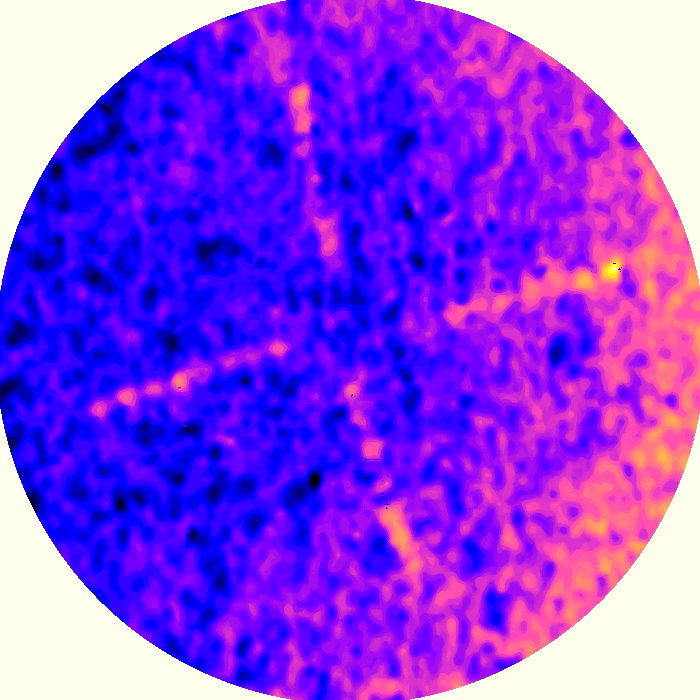}\hfill
  \includegraphics[width=0.26\columnwidth]{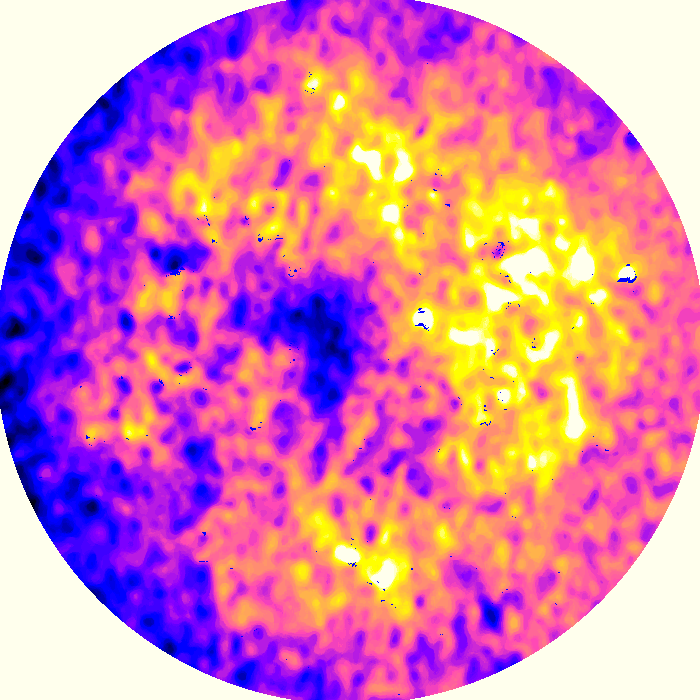}\hfill
  \includegraphics[width=0.26\columnwidth]{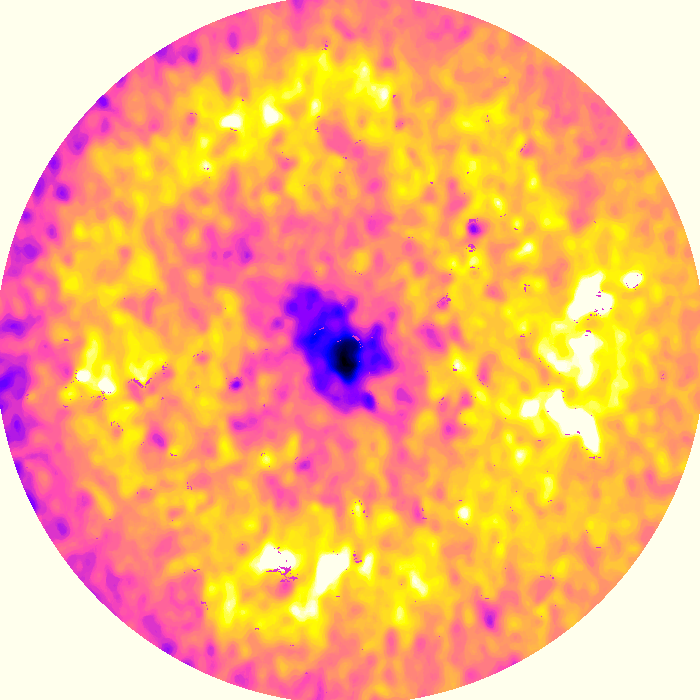}\hfill
 \includegraphics[width=0.1\columnwidth]{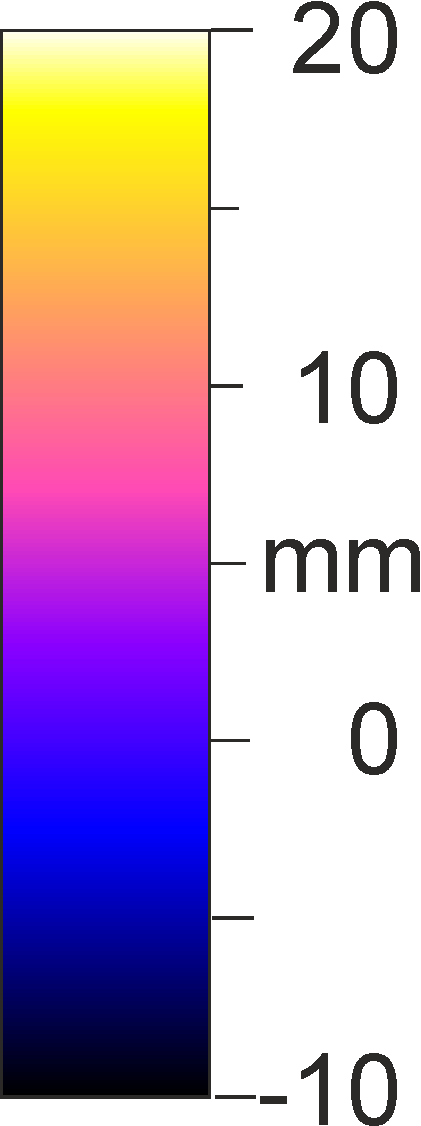}
  \caption{Height profile of the complete surface of the granular bed. The heights are shown for the initial,
  flat surface (left), and after 5 (middle) and 15 (right) revolutions of the bottom plate. The height difference to the nominal
  fill height of 100 mm is color-coded.}
  \label{fig:snapshots}
\end{figure}

Analogous to the earlier findings for rods, where heaping in the center and outward surface flow was found, the radially inward surface flow of
crosses is related to a \rs{change of the surface profile} in the rotation center. We show examples of height profiles recorded with the stereo camera in Fig.~\ref{fig:snapshots}. Initially, the granular bed is flat within $\approx 5$~mm \rs{(approximately half the cross size). The tracer crosses that were placed on the flat granular bed can be identified as four bright lines. They sink to the surrounding level within less than one rotation.}
After a few rotations, the granular bed above the shear zone expands. This is indicated by a height change of about 5\% to 10\% as the effect of Reynolds dilatancy \cite{Reynolds1885}. The material cannot expand in the plane quickly, so the local surface is elevated. After 15 rotations, this effect is even stronger, and the elevation shifts
outward. Remarkably, a dimple is formed in the middle. This region has a diameter of roughly 3 to 4 cm.
Marked crosses that are dragged into this sink disappear from the surface.

\begin{figure}[htbp]
   \includegraphics[width=\columnwidth]{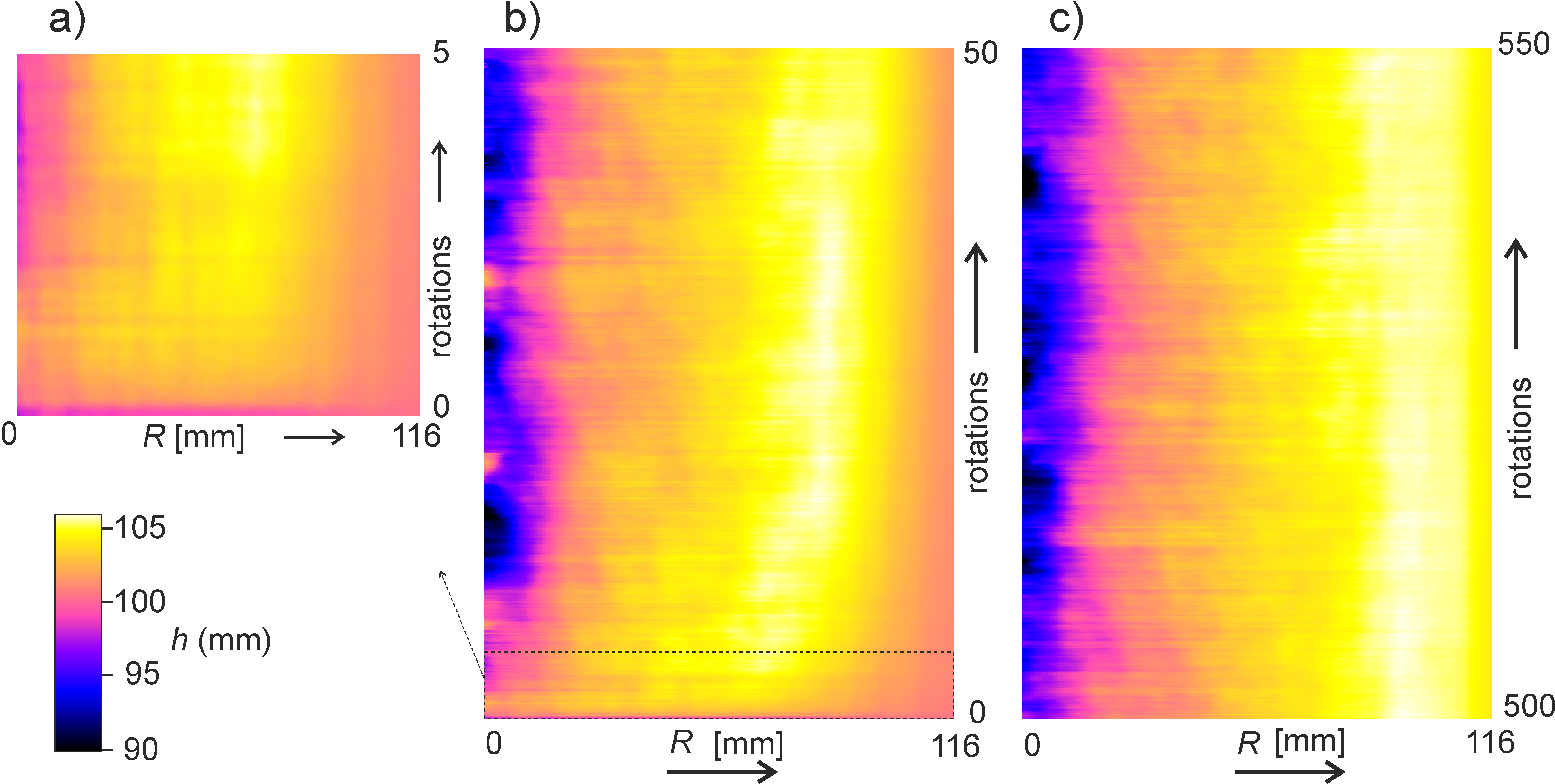}
  \caption{Space-time plots of the azimuthally averaged surface profiles \rs{ (local bed heights $h$), $H=100~{\rm mm} =0.8~R_0$, $\omega_0=6$ rpm.
 a) visualizes the bed expansion due to Reynolds dilatancy within the first 5 disk rotations, b) shows the evolution of the dimple, and c) reflects the situation after a stationary profile is established.}}
  \label{fig:stp}
\end{figure}

\rs{Figure \ref{fig:stp} shows typical space-time plots of the surface profile, averaged over regions of equal $R$. In Fig. \ref{fig:stp}a, the expansion of the bed above the sheared region due to Reynolds dilatancy is seen. In Fig. \ref{fig:stp}b, the dimple forms and reaches its final depth within roughly 10 to 20 rotations. At the same time, the elevated region near
$R_0/2$ moves outward and expands to some saturation value. The stationary state is reflected in Fig. \ref{fig:stp}c. Note that the depth of the sink fluctuates considerably. One obvious reason is that near $R=0$, only a small surface area contributes to the azimuthally averaged statistics. Similar local fluctuations are averaged out in the data for larger distances $R$.}

\begin{figure}[ht]
   \includegraphics[width=0.8\columnwidth]{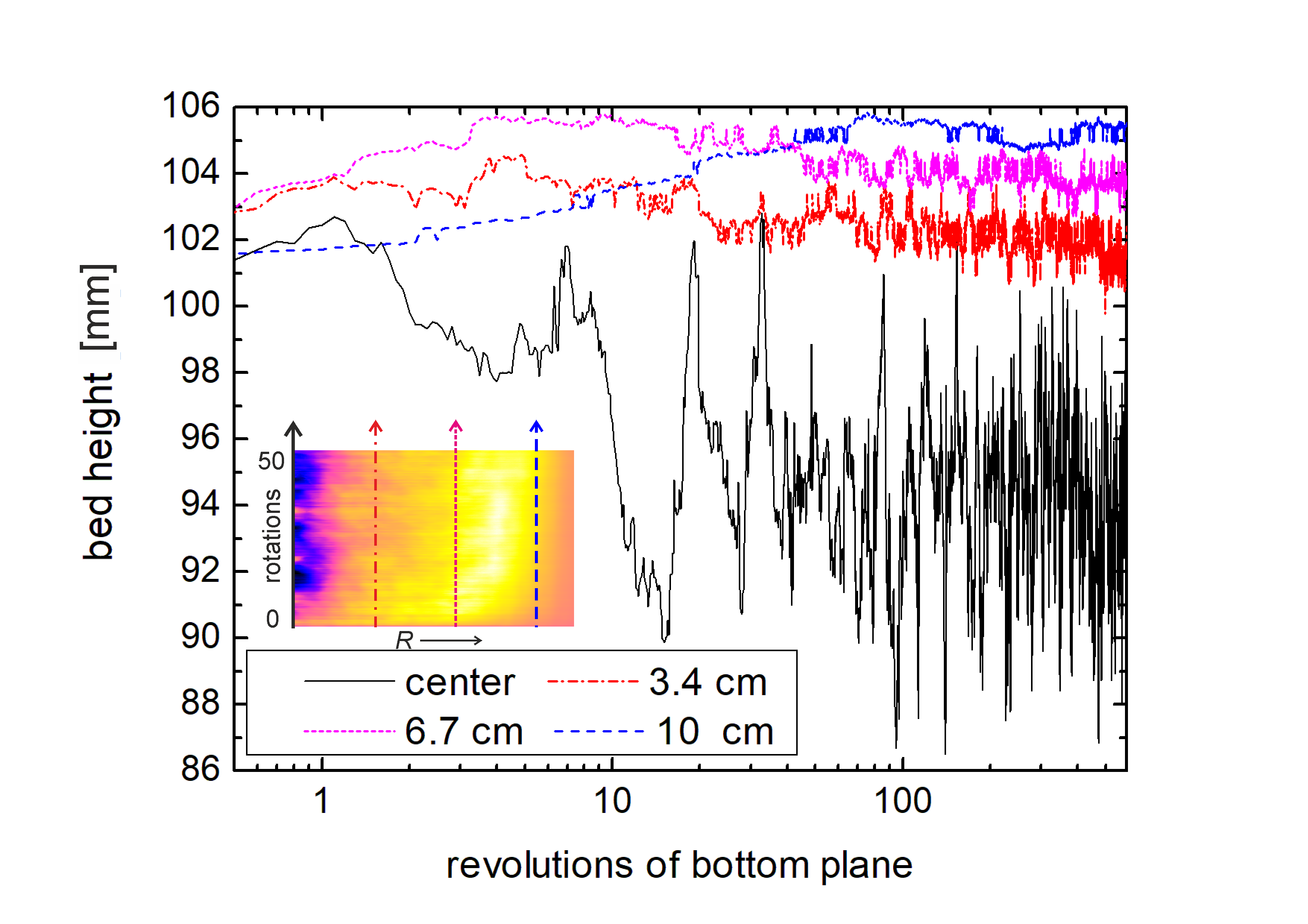}\\
  \caption{\rs{Azimuthally averaged time dependence of the bed heights at three selected radii $0.266~R_0$, $0.533~R_0$, $0.8~R_0$, and in the center.
   The inserted space-time plot is that of Fig.~\ref{fig:stp}b, where the positions of the four graphs are marked. Note the logarithmic abscissa.}   $H\approx 0.8~R_0$, $\omega_0=6$ rpm.
  }
  \label{fig:timeplots}
\end{figure}

A more quantitative evaluation is presented in Fig.~\ref{fig:timeplots}. Four typical radii were selected for plots of the time dependent bed heights. The central curve is averaged over a circular region with about 1 mm diameter, all other graphs are averages over all points with the given radial distance ($\approx 0.33~ {\rm mm}\times 2\pi R$). Since the initial changes are more dramatic than the later dynamics, we have chosen a logarithmic time axis.

The graphs start at roughly the same level, then the central height drops within the first 10 revolutions of the bottom disk (corresponding to about 1/4 rotation of the central surface).
Within the first 2 to 4 rotations of the bottom disk, the bed expands by about 5\% in the range $R\approx 0.3~R_0$ to 0.6~$R_0$. In the outer region ($R\approx R_0$) the expansion occurs later, within the first 30 to 50 bottom disk rotations. While that region finally elevates by approximately 5 \%, the inner zones slightly collapse again. Within the fluctuations seen in the graphs, the profile becomes stationary, while the convective flow is maintained.
\rs{It is interesting to mention in passing that a reversal of the rotation sense of the bottom disk leads to an immediate collapse of the granular bed height within less than 1/4 bottom disk rotation.}

We have performed similar experiments with thinner spatial crosses of the same volume and an aspect ratio $\rho=4$ ($\ell=14.8$~mm,
$d=3.7$~mm). In these experiments,
the same features as for the thicker $\rho=3$ crosses are confirmed qualitatively. Dimple formation as well as surface flow are found under
comparable geometrical conditions. Similar experiments performed with crosses of aspect ratio $\rho=6$, however, show no noticeable convective flow nor a central dimple in the same range of geometrical parameters. It is possible that
at other fill levels and on much longer time scales (much more revolutions) these effects might be recognized.
The central surface rotation rate $\omega_c$ is shown in Fig.~\ref{fig:tangentialFlow}. At comparable fill heights, it is much slower than for the $\rho=3$ crosses. Regarding the level of convexity \cite{Azema2013}, it is intuitively clear that this feature is larger for the $\rho=3$ crosses than for $\rho=6$. Thus, it is somewhat counterintuitive, if one attributes the observed phenomena to the nonconvex shapes of the spatial crosses, that the effect is stronger for the thicker crosses with $\rho=3$.

A physical explanation of the observed phenomena is still pending. Possibly, the reason is related to the strong Reynolds dilatancy of the crosses.
This effect is much smaller for rod-like particles where it is at least partially compensated by a compaction through shear-alignment \cite{Wegner2014}.
It may be easier for a spatial cross in the depth of the granular bed to enter the shear zone from the side because of the lower packing fraction.
This would lead to an absorption of crosses by the shear zone, related to a subsurface net flow into the shear zone. Such a flow would create voids
in the center of the container where crosses are pulled down, leaving a dimple and creating an inward surface flow.
A better understanding requires a more comprehensive investigation of this phenomenon, including non-invasive 3D
imaging with X-ray Computed Tomography or Nuclear Magnetic Resonance. In addition, the development of numerically efficient and accurate DEM simulation of the system under consideration would be highly desirable.

This study received funding from the German Science Foundation within project STA 425/46-1 and from the European Union's Horizon 2020 research and innovation programme under the Marie Sk\l{}odowska-Curie
grant agreement {\sc CALIPER} No 812638. \rs{We are particularly indebted to Joshua Dijksman for providing the spatial crosses.}

%\bibstyle{prbst}
%\bibliography{hexapods}

%merlin.mbs apsrev4-1.bst 2010-07-25 4.21a (PWD, AO, DPC) hacked
%Control: key (0)
%Control: author (8) initials jnrlst
%Control: editor formatted (1) identically to author
%Control: production of article title (-1) disabled
%Control: page (0) single
%Control: year (1) truncated
%Control: production of eprint (0) enabled
%

\end{document}